\documentclass[aps,prc,twocolumn,groupaddress,showkeys]{revtex4-1}
\usepackage{graphicx} 
\usepackage{dcolumn} 
\usepackage{bm}
\usepackage{color}
\usepackage{hyperref}

\begin{document}

\title{Probing the density dependence of the symmetry energy by nucleon flow}
\author{Xiao-Hua Fan$^{1,2,3}$}
\author{Gao-Chan Yong$^{1,3}$}
\email[]{yonggaochan@impcas.ac.cn}
\author{Wei Zuo$^{1,3}$}
\affiliation{$^1$Institute of Modern Physics, Chinese Academy of Sciences, Lanzhou 730000, China\\
$^2$School of Physical Science and Technology, Lanzhou University, Lanzhou 730000, China\\
$^3$School of Nuclear Science and Technology, University of Chinese Academy of Sciences, Beijing 100049, China}

\begin{abstract}

In the framework of the isospin-dependent Boltzmann-Uehling-Uhlenbeck transport model,
sensitive regions of some nucleon observables to the nuclear symmetry energy are studied. It is found that the symmetry energy sensitive observable n/p ratio in the $^{132}$Sn+$^{124}$Sn reaction at 0.3 GeV/nucleon in fact just probes the density-dependent symmetry energy below the density of $1.5\rho_0$ and effectively probes the density-dependent symmetry energy around or somewhat below the saturation density. Nucleon elliptic flow can probe the symmetry energy from the low-density region to the high-density region when changing the incident beam energies from 0.3 to 0.6 GeV/nucleon in the semi-central $^{132}$Sn+$^{124}$Sn reaction. And nucleon transverse and elliptic flows in the semi-central $^{197}$Au+$^{197}$Au reaction at 0.6 GeV/nucleon are more sensitive to the high-density behavior of the nuclear symmetry energy. One thus concludes that nucleon observables in the heavy reaction system and with higher incident beam energy are more suitable to be used to probe the high-density behavior of the symmetry energy. The present study may help one to get more specific information about the density-dependent symmetry energy from nucleon flow observable in heavy-ion collisions at intermediate energies.


\end{abstract}

\maketitle

\section{INTRODUCTION}

The density-dependent symmetry energy, which characterizes the isospin-dependent part of nuclear equation of state (EOS), is a hot topic owing to its importance in both nuclear physics and astrophysics \cite{Tsang,Neutron Radii,Supernova1,science}. In nuclear physics, the symmetry energy is a key factor in understanding many issues such as the nuclear bound energy, deformation, the density and radius of neutron distribution, the stability of super-heavy nuclei, and the neutron skin thickness of neutron-rich nuclei \cite{wang,li1,dong1}. Also, as the enormous high isospin asymmetric nuclear matter was produced in the evolution of the cosmos, the symmetry energy plays a crucial role in answering a variety of questions in astrophysics, like the structures, composition, and cooling of neutron stars, and the birth and core-collapse evolution of supernovas \cite{Astron,supernovae2}. There is a noteworthy fact that the symmetry energy and its density dependence determine whether the direct Urca process is allowed or not in neutron stars \cite{URCA1,URCA2}.

To pin down the symmetry energy in the broad density region, people have made great efforts, including theoretical researches \cite{zuo1,OR,fan,li2} and experimental studies \cite{betty09,fopi16,xiao17}, both terrestrial \cite{XIAO,ex1,ex2} and celestial \cite{CE1,CE2}. Fortunately, the symmetry energy and the slope parameter governing its density-dependence around the saturation density of the nuclear matter have been roughly constrained \cite{top14,Chen}. However, due to the complexity of the nuclear force and nuclear many-body problem, the
high-density behavior of the symmetry energy given by various theoretical models is still very much controversial \cite{guo2013,guo2014}. It is thus necessary to find a way to meet this challenge. Nowadays, heavy-ion collisions with rare isotopes can be used to probe the nuclear symmetry energy, especially at high densities. At present, many observables, which have been identified to be sensitive to the symmetry energy, like the free neutron to proton ratio n/p \cite{li97,np1,np2}, the $\pi^+/\pi^-$ ratio \cite{pai1,pai2,pai3,pai4}, radial flow \cite{li3}, isospin fraction \cite{ISOF1,ISOF2}, isospin diffusion \cite{TSANG2,CHEN2}, and transverse and elliptic flows \cite{YONG1}, could be effective probes of the symmetry energy.

{\color{blue}{However, it is not straightforward that one can obtain the high-density behavior of the symmetry energy when heavy-ion collisions produce dense matter with densities well above saturation density}}. One needs to study in detail which density region that the symmetry energy sensitive observables probed. Allowing for the fact that the isotope reaction reaction $^{132}$Sn+$^{124}$Sn at a beam energy of 0.3 GeV/nucleon is being carried out at RIKEN in Japan \cite{japan1,japan2},
and the $^{197}$Au+$^{197}$Au reaction at 0.6 GeV/nucleon is being planned at FOPI/GSI and CSR/Lanzhou \cite{fopi16,csr}, in this study, we try to give the specific density region for which some nucleon observables are mainly probed. The present work may help one to constrain the high-density behavior of the symmetry energy more effectively.

\section{METHODS AND RESULTS}

The present work is based on the framework of the isospin-dependent Boltzmann-Uehling-Uhlenbeck (BUU) transport model. The used Skyrme-type parametrization for the isoscalar term of the mean field potential \cite{U}
gives the ground-state compressibility coefficient of nuclear matter $K$ = 230 MeV \cite{YONG2}. Similar to the studies in the Refs. \cite{pai4,LIU1,LIU2}, for the purpose of determining in which density region the observables are sensitive to the symmetry energy, a specific density-dependent symmetry energy form $E_{sym}= 32(\rho/\rho_0)^{\gamma= 2}$ is chosen in a certain density region. The choice of the symmetry energy parameter $\gamma$ is just for the convenience of research. To gain the relative sensitivity of the symmetry energy sensitive observables in different density regions, the symmetry energy from zero to maximum density is added step by step in the transport model to simulate the heavy-ion collision. Five various cases are considered; i.e., the symmetry energy is added in the density regions $0-0.5\rho_0$, $0- \rho_0$, $0-1.5\rho_0$, and $0-2\rho_0$, and the case of no symmetry energy is applied in the whole density region.

\begin{figure}
\includegraphics[width=8 cm]{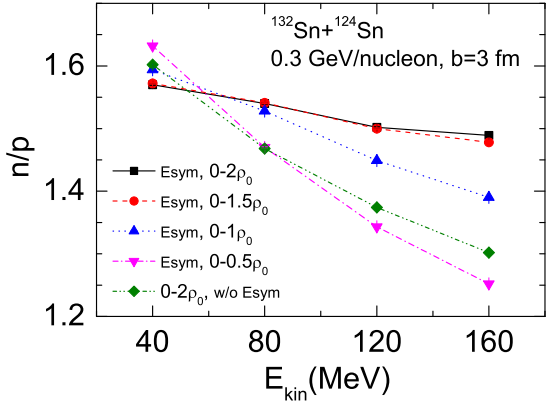}
\caption{\label{fig:rnpsn300} (Color online) The relative sensitivity of the symmetry energy sensitive observable free neutron to proton ratio n/p as a function of kinetic energy in the central $^{132}$Sn+$^{124}$Sn reaction at 0.3 GeV/nucleon.}
\end{figure}
Figure \ref{fig:rnpsn300} shows the relative sensitivity of the symmetry energy sensitive observable free neutron to proton ratio n/p as a function of kinetic energy in the $^{132}$Sn+$^{124}$Sn reaction at a beam energy of 0.3 GeV/nucleon and an impact parameter of 3 fm. Since protons are pushed to be away from the center of the reaction system by the Coulomb potential, it is thus naturally seen that the n/p ratio in each case decreases as the kinetic energy increases, especially without the action of the symmetry energy.
The fact that the nuclear symmetry potential repels neutrons and attracts protons causes the n/p ratio at high kinetic energies to increase step by step by adding the symmetry energy step by step except in the density region of $0-0.5\rho_0$.
Since the symmetry energy in the density region $0-0.5\rho_0$ repels neutrons from the low-density region $0-0.5\rho_0$ to relative high-density region, one thus sees the free n/p ratio decrease with the symmetry energy in the density region $0-0.5\rho_0$ due to the anti-repulsive effects of the symmetry energy at very low densities.
The effect of the symmetry energy in the density region $1.5\rho_0-2\rho_0$ is negligible
since the n/p ratios almost overlap when adding the symmetry energy in the density regions $0-1.5\rho_0$ and $0-2\rho_0$.
The effects of the symmetry energy shown in the region $0.5\rho_0-\rho_0$ are somewhat larger than that shown in the region $\rho_0-1.5\rho_0$.
From Fig. \ref{fig:rnpsn300}, it is seen that the symmetry energy sensitive observable n/p ratio in the $^{132}$Sn+$^{124}$Sn reaction at 0.3 GeV/nucleon in fact just probes the density-dependent symmetry energy below the density of $1.5\rho_0$ and effectively probes the density-dependent symmetry energy around or somewhat below the saturation density.

Nucleon directed and elliptic flows in
heavy-ion collisions can be derived from the
Fourier expansion of the azimuthal distribution
\cite{Voloshin,dan1998,wang14,flow}, i.e.,
\begin{equation}
\frac{dN}{d\phi}\propto1+2
\displaystyle{\sum_{i=1}^{n}}v_{n}\cos(n\phi).
\end{equation}
The nucleon elliptic flow $v_{2}$ can be obtained from
\begin{equation}
v_{2}=\langle\cos(2\phi)\rangle=\langle\frac{p_{x}^{2}-p_{y}^{2}}{p_{t}^{2}}\rangle
\end{equation}
and can be used to probe the nuclear symmetry energy \cite{flow}.
\begin{figure}
\includegraphics[width=8 cm]{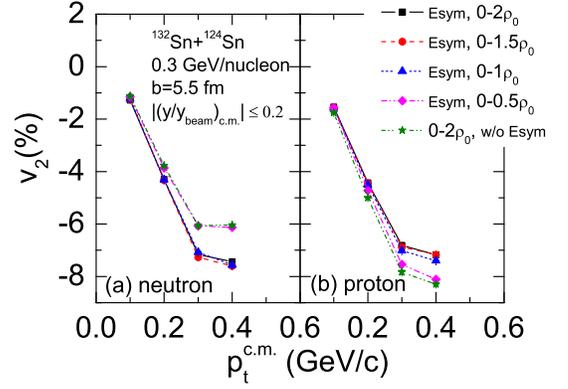}
\caption{\label{fig:v2sn3002} (Color online) The relative sensitivity of neutron and proton elliptic flows as a function of transverse momentum in the semi-central $^{132}$Sn+$^{124}$Sn reaction at 0.3 GeV/nucleon.}
\end{figure}
The neutron and proton elliptic flows as a function of transverse momentum in the $^{132}$Sn+$^{124}$Sn reaction at an incident energy of 0.3 GeV/nucleon and an impact parameter of 5.5 fm are plotted in Fig. \ref{fig:v2sn3002}.
The effects of the symmetry energy shown in Fig. \ref{fig:v2sn3002}(a) are roughly equal when adding the symmetry energy in the density regions $0-\rho_0$, $0-1.5\rho_0$ and $0-2\rho_0$, respectively. The neutron elliptic flow without adding the symmetry energy is almost the same as the one when adding the symmetry energy in the density region $0-0.5\rho_0$. That is to say, only the effect of the symmetry energy in the density region $0.5\rho_0-\rho_0$ is distinguishable.
Thus, it suggests that the neutron elliptic flow in the $^{132}$Sn+$^{124}$Sn reaction at 0.3 GeV/nucleon mainly probes the density-dependent symmetry energy in the region  $0.5\rho_0-\rho_0$, which is similar to the proton elliptic flow shown in
of Fig. \ref{fig:v2sn3002}(b). The neutron and proton elliptic flows in the $^{132}$Sn+$^{124}$Sn reaction at the incident energy of 0.3 GeV/nucleon thus cannot be used to probe the high-density behavior of the symmetry energy.

\begin{figure}
\includegraphics[width=8 cm]{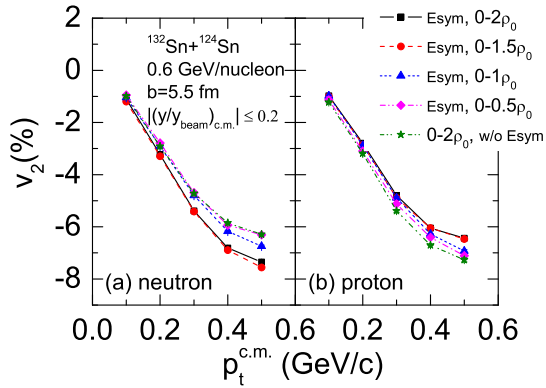}
\caption{\label{fig:v2sn6002} (Color online) The relative sensitivity of neutron and proton elliptic flows as a function of transverse momentum in the semi-central $^{132}$Sn+$^{124}$Sn reaction at 0.6 GeV/nucleon.}
\end{figure}
To probe the density-dependent symmetry energy at higher densities, we change the incident beam energy from 0.3 to 0.6 GeV/nucleon.
The relative sensitivity of neutron and proton elliptic flows as a function of transverse momentum in the $^{132}$Sn+$^{124}$Sn reaction at the incident beam energy of 0.6 GeV/nucleon is shown in Fig. \ref{fig:v2sn6002}.
Comparing with the situation shown in Fig. \ref{fig:v2sn3002}, the effects of the symmetry energy on the neutron elliptic flow become larger in the density region $\rho_0-1.5\rho_0$. And also the effects of the symmetry energy on the proton elliptic flow become more evident in this density region. It is concluded that nucleon elliptic flow can probe the symmetry energy from the
low-density region to high-density region when changing the incident beam energies from 0.3 to 0.6 GeV/nucleon in the $^{132}$Sn+$^{124}$Sn reaction with an impact parameter of 5.5 fm.

\begin{figure}
\includegraphics[width=8 cm]{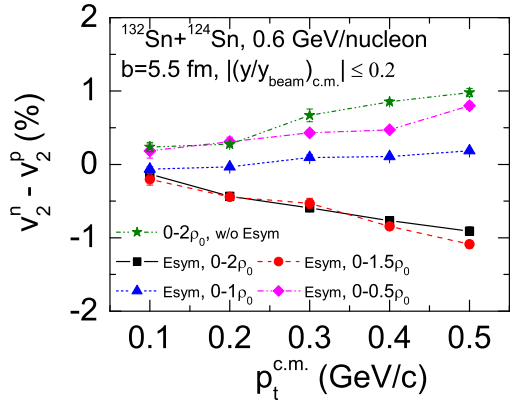}
\caption{\label{fig:v2sn600} (Color online) The relative sensitivity of the difference between neutron elliptic flow and proton elliptic flow as a function of transverse momentum in the semi-central  $^{132}$Sn+$^{124}$Sn reaction at 0.6 GeV/nucleon.}
\end{figure}
Since the difference between neutron and proton elliptic flows is more useful to probe the symmetry energy \cite{flow}, shown in Fig. \ref{fig:v2sn600} is the difference between neutron elliptic flow and proton elliptic flow as a function of transverse momentum in the $^{132}$Sn+$^{124}$Sn reaction at 0.6 GeV/nucleon. It is seen that the effects of the symmetry energy in the density region $\rho_0-1.5\rho_0$ are somewhat larger than in the region $0-\rho_0$. However, the difference between neutron and proton elliptic flows in a medium nuclei $^{132}$Sn+$^{124}$Sn collision at 0.6 GeV/nucleon still cannot probe the symmetry energy in the region $1.5\rho_0-2\rho_0$.

\begin{figure}
\includegraphics[width=8 cm]{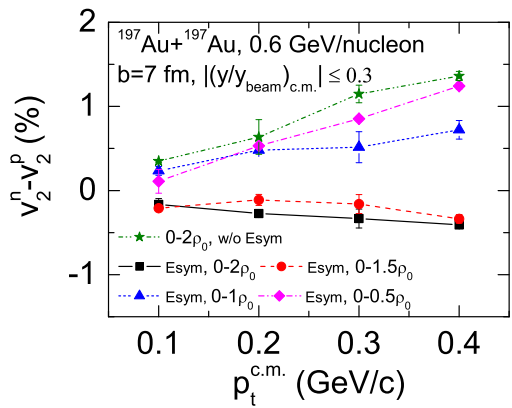}
\caption{\label{fig:v2au600} (Color online) The relative sensitivity of the difference between neutron elliptic flow and proton elliptic flow as a function of transverse momentum in the semi-central $^{197}$Au+$^{197}$Au reaction at 0.6 GeV/nucleon.}
\end{figure}
Since the medium nuclei $^{132}$Sn+$^{124}$Sn collision at 0.6 GeV/nucleon cannot effectively probe the symmetry energy at high densities, and in heavy nuclei collisions the symmetry potential may act on nucleons longer than in light nuclei collisions, in Fig. \ref{fig:v2au600} we show the relative sensitivity of the difference between neutron elliptic flow and proton elliptic flow in the heavier nuclei semi-central collision $^{197}$Au+$^{197}$Au at a beam energy of 0.6 GeV/nucleon.
It is seen that the effects of the symmetry energy in the density region $\rho_0-1.5\rho_0$ are evidently larger than in the region $0-\rho_0$. Unfortunately, the difference between neutron and proton elliptic flows in the $^{197}$Au+$^{197}$Au collision at 0.6 GeV/nucleon still cannot probe the symmetry energy at densities above $1.5\rho_0$. Nevertheless, a heavy reaction system at higher incident beam energy is more suitable to be used to probe the high-density behavior of the symmetry energy.

\begin{figure}
\includegraphics[width=9 cm]{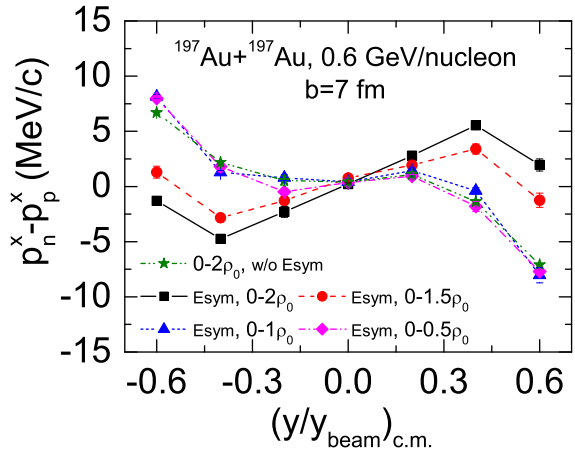}
\caption{\label{fig:pxnpAu600} (Color online) The relative sensitivity of the difference between neutron transverse flow and proton transverse flow as a function of rapidity in the $^{197}$Au+$^{197}$Au reaction at 0.6 GeV/nucleon.}
\end{figure}
Inspired by the above result, as an alternative symmetry energy sensitive observable, shown in Fig. \ref{fig:pxnpAu600}, we also studied the difference between neutron transverse flow and proton transverse flow as a function of rapidity in the $^{197}$Au+$^{197}$Au reaction at 0.6 GeV/nucleon. The difference between neutron transverse flow and proton transverse flow as a function of rapidity reads
\begin{eqnarray}
p_{n}^{x}-p_{p}^{x} &=
&\frac{1}{N_{n}(y)}\sum_{i=1}^{N_{n}(y)}p_{i}^{x}(y)-\frac{1}{N_{p}(y)}\sum_{i=1}^{N_{p}(y)}p_{i}^{x}(y),
\end{eqnarray}%
where $N_{n}(y)$ and $N_{p}(y)$ are the numbers of neutrons and protons, respectively, at rapidity $y$; $p_{i}^{x}(y)$ is the transverse momentum of the free nucleon at
rapidity $y$. From Fig. \ref{fig:pxnpAu600}, it is seen that the difference between neutron transverse flow and proton transverse flow in the semi-central $^{197}$Au+$^{197}$Au reaction at 0.6 GeV/nucleon mainly probes the symmetry energy in the density region $\rho_0-1.5\rho_0$, while it is not sensitive to the symmetry energy below the saturation density. It is also seen that this observable in such a reaction is also sensitive to the symmetry energy in the density region $1.5\rho_0-2\rho_0$ to some extent. Comparing the results shown in Fig. \ref{fig:pxnpAu600} with the results in Fig. \ref{fig:v2au600}, it is concluded that the difference between neutron and proton transverse flows and the difference between neutron and proton elliptic flows are more suitable to probe the high-density behavior of the nuclear symmetry energy in the semi-central $^{197}$Au+$^{197}$Au reaction at 0.6 GeV/nucleon.

In Ref.~\cite{fopi16}, the sensitivity of elliptical flow ratio to the various density regimes probed in heavy-ion collisions was similarly studied quantitatively by switching the density dependent symmetry energy at certain density regions. For Au + Au reactions at 0.4 GeV/nucleon, it is observed that the maximum sensitivity of the neutron/proton elliptic
flow ratio lies in the 1.4-1.5$\rho_0$ region. The results obtained in Ref.~\cite{fopi16} are thus similar to our results in the present studies. And also the sensitivity of the symmetry-sensitive-observable decreases with the increase of incident beam energy;
nucleon flow in heavy-ion collisions thus cannot effectively probe the symmetry energy above 1.5$\rho_0$. Alternatively, to constrain the symmetry energy above 1.5$\rho_0$ by nucleon flow, one has to constrain the symmetry energy below 1.5$\rho_0$.

The effects of short range correlations of nucleons, especially the high-momentum tail of the nucleon momentum distribution in nuclei, would play a role in symmetry-energy-sensitive observables in heavy-ion collisions \cite{src1,src2,src3,src4}. The effects of nucleon short range correlations
on nucleon transverse and elliptic flows are shown in Ref.~\cite{src3}. Although the short range correlations may affect the sensitivities of observables to the symmetry energy, they may not change the relative sensitivity of an observable to the symmetry energy in different density regions. This is because the short range correlations are less affected by the density of nuclear matter as well as by the tiny changes of asymmetry of nuclear matter in heavy-ion collisions.
In this case, the physical results of the present study would not change much with or without short range correlations.

Since the present studied nucleon elliptic or transverse flows are both as a function of momentum or rapidity, the momentum dependencies of the symmetry potential at different densities would more or less affect the relative sensitivity of an observable to the symmetry energy in different density regions \cite{ditoro}. This question deserves further study.

\section{Conclusions}

Within the isospin-dependent BUU transport model, we studied the regions of some nucleon observables that are sensitive to the nuclear symmetry energy. It is found that nucleon observables in the $^{132}$Sn+$^{124}$Sn reaction at 0.3 GeV/nucleon cannot effectively probe the density-dependent symmetry energy above a density of $1.5\rho_0$. However, nucleon observables can probe the symmetry energy from the low-density region to high-density region when changing the incident beam energies from 0.3 to 0.6 GeV/nucleon, especially in the heavy reaction system.

{\color{blue}{One generally considers that the symmetry energy sensitive observable would probe the symmetry energy at high densities if the maximum compression density reached in heavy-ion collisions is evidently above the saturation density. In fact, this question is not straightforward, not only because the symmetry energy sensitive observable in heavy-ion collisions generally does not reflect the value of the symmetry energy but rather its slope, but also the symmetry energy sensitive observable usually suffers interference from the low-density nuclear matter.}}

Presently there is no good way to study the density region that the symmetry energy sensitive observable probed, except for the unphysical operation of changing the symmetry energy in a certain density region. Therefore, some other methods of extracting the specific density region that the symmetry energy sensitive observable probed are still needed to cross-check our obtained physical results, and not making the deduction merely from imagination.

\section{Acknowledgements}

This work is supported in part by the National Natural Science
Foundation of China under Grant Nos. 11775275, 11435014 and the 973 Program of China (2007CB815004).

\end{document}